\documentclass[aps,prl,prabib,twocolumn,showpacs,nofootinbib]{revtex4}
\usepackage{graphicx} \usepackage{amsmath} \usepackage{amssymb}
\usepackage{amsfonts} \usepackage{bm}

\begin{document}

\newcommand{\be}{\begin{equation}} \newcommand{\ee}{\end{equation}}
\newcommand{\bea}{\begin{eqnarray}}\newcommand{\eea}{\end{eqnarray}}


\title{Electron capture and scaling anomaly in polar molecules}

\author{Pulak Ranjan
Giri} \email{pulakranjan.giri@saha.ac.in}
\author{Kumar S. Gupta} \email{kumars.gupta@saha.ac.in}

\affiliation{Theory Division, Saha Institute of Nuclear Physics, 1/AF Bidhannagar, Calcutta
700064, India} 

\author{S. Meljanac} \email{meljanac@irb.hr}
\author{A. Samsarov} \email{asamsarov@irb.hr}

\affiliation{Rudjer Bo\v{s}kovi\'c Institute, Bijeni\v cka  c.54, HR-10002
Zagreb, Croatia}

\begin{abstract}
We present a new analysis of the electron capture mechanism in polar 
molecules, based on von Neumann's theory of self-adjoint extensions. Our 
analysis suggests that it is theoretically possible for polar molecules to 
form bound states with electrons, even with dipole moments smaller than the 
critical value $D_0$ given by $1.63\times10^{-18}$ esu cm. This prediction 
is consistent with the observed anomalous electron scattering in $H_2S$ and 
$HCl$, whose dipole moments are smaller than the critical value $D_0$.  We 
also show that for a polar molecule with dipole moment less than $D_0$, 
typically there is only a single bound state, which is in qualitative 
agreement with observations. We argue that the quantum mechanical scaling 
anomaly is responsible for the formation of these bound states. 

\end{abstract}



\date{\today}

\maketitle

The experimentally observed anomalous scattering of electrons by a class of polar molecules \cite{hurst1,hurst2,hurst3} is often attributed to the electron capture by the dipole field of the polar molecules \cite{turner1,turner2,turner3}. It is usually assumed that a dipole bound state of an electron is possible only if the coefficient of the inverse square interaction term is sufficiently negative, leading to a ``fall to the centre" condition \cite{landau}. The point dipole model of the polar molecule predicts that the critical dipole moment \cite{fermi} $D_0$ necessary for the electron capture has a value $D_0 = 1.63\times10^{-18}$ esu cm \cite{leblond,turner,garret,camblong1}, which continues to be valid for an extended dipole as well \cite{leblond}. The captured electrons are usually weakly bound, which makes such bound states hard to detect unless the dipole moments are large compared to $D_0$. Consequently, most of the experiments are performed with molecules having large dipole moments, for which the above value of $D_0$ is consistent with the experimental data. 

However, there are certain polar molecules, e.g. $H_2S$ and $HCl$ which have dipole moments less than $D_0$ and yet exhibit anomalous electron scattering and can capture electrons \cite{leblond,rohr}. The simple model of the inverse square interaction which predicts the above value of $D_0$, at the first sight, seems to be inconsistent with the data for $H_2S$ and $HCl$. Moreover, the same theoretical model predicts an infinite number of bound states for  dipole bound anions, whereas most experiments observe only a single bound state \cite{amy}. 

We may think that these inconsistencies arise from the fact that the pure dipole model of a polar molecule ignores various short range interactions that play a role in the electron capture. However, the inverse square interaction between the electron and the polar molecule encodes the main aspects of the long distance dynamics leading to the formation of weakly bound states \cite{garret}. Within this approximation, and without using any detailed knowledge of the short distance interactions, it is reasonable to represent their effects through the boundary conditions obeyed by the Hamiltonian. Note that the molecular forces are usually not dissipative, and the corresponding Hamiltonian remains always self-adjoint \cite{reed}. Thus, the problem now reduces to finding all possible boundary conditions for a quantum system with inverse square interaction, such that the Hamiltonian is self-adjoint.  This is usually achieved by using von Neumann's theory \cite{reed}  of self-adjoint extensions, which provides all the possible boundary conditions such that the Hamiltonian maintains self-adjointness. The theory of self-adjoint extension of the inverse square potential \cite{meetz} has a long and interesting list of applications ranging from microscopic physics to black holes \cite{biru,kumar1,kumar2,kumar3,bh,stjep}, and we shall apply this well established technique to explain the electron capture by polar molecules. Our approach is based on a detailed quantum mechanical treatment of the system, which suggests the possibility of the existence of bound states even though the dipole moment is less than $D_0$. This analysis, which predicts a lower value of the critical dipole moment is consistent with all experimental data including that for $H_2S$ and $HCl$. It also predicts the theoretical possibility of electron capture by a much larger class of polar molecules. Another interesting feature of our analysis is that when the dipole moment is less than $D_0$, the system admits only a single bound state, which is in qualitative agreement with the experimental data \cite{amy}. It should be noted that our analysis presented below is not applicable to molecules with dipole moments greater than $D_0$, for which the existing treatment in the literature is consistent with the experimental data.

The simple quantum mechanical model of inverse square interaction does not contain any dimensionful parameter. It has been argued that such a system exhibits quantum mechanical scaling anomaly \cite{camblong1}, leading to the formation of bound states. It is known that the self-adjoint extensions can lead to anomalies \cite{esteve}, which in this case allows the polar molecule to capture an electron. We shall show that the scaling anomaly exists even for systems with dipole moments less than $D_0$, thus allowing a much larger class of polar molecules to form bound states with electrons.

Consider an electron of charge $e$ and mass $\mu$ moving in a
point dipole field with dipole moment $D$. We take the $z$ axis along the
dipole moment. Schr\"{o}dinger equation for the electron in the spherical
polar coordinates  $(r,\theta,\phi)$ can be written as
\begin{equation}
\left[-\frac{\hbar^2}{2\mu}\nabla^2 + \frac{eD}{r^2}\cos\theta\right]\Psi=
E\Psi\,,
\label{schrodinger}
\end{equation}
where $E$ is the energy.
In these coordinates, the wavefunction can be separated as
$\Psi(r,\theta,\phi) =  \frac{1}{r}R(r)\Theta(\theta)e^{im\phi}$, which leads to 
the radial equation 
\be H_r R(r)
\equiv  \left [ -\frac{d^2}{dr^2} +\frac{\lambda}{r^2} \right ]R(r)
= \epsilon R(r)\,, \label{radial} \ee where $H_r$ is the radial
Hamiltonian, $\epsilon = \frac{2\mu E}{\hbar^2}$ is the eigenvalue of the radial eqn. (\ref{radial}) and $\lambda$ is the eigenvalue of the angular equation,
given to $\mathcal{O}((\frac{1}{6}d^2)^3)$ by \cite{leblond}
\begin{equation}
\lambda =-\frac{1}{6}d^2 + \frac{11}{1080}d^4 - \frac{133}{97200}d^6 +
... \,,
\label{coupling}
\end{equation}
with ${d= \frac{2\mu e D}{\hbar^2}}$. It is usually assumed in the literature that in order for the operator $H_r$ in eqn. (\ref{radial}) to admit
a bound state, the ``fall to the centre" condition \cite{landau} given by $\lambda < - \frac{1}{4}$ must be satisfied \cite{leblond}. 
This assumption leads to the critical dipole moment $D\geq D_0 =
1.63\times10^{-18}$ esu cm. Using von Neumann's approach, and with a very general assumption that the Hamiltonian is self-adjoint, we shall now show that it is possible to form bound states in this system with a weaker condition on $\lambda$ and correspondingly for smaller values of $D_0$.

The Hamiltonian $H_r$ is a real symmetric (Hermitian)
operator on the domain $D(H_r) \equiv \{\phi (0) = \phi^{\prime} (0) = 0,~
\phi,~ \phi^{\prime}~  {\rm absolutely~ continuous} \} $.  Following von
Neumann's method \cite{reed}, in order to determine whether  $H_r $ is self-adjoint in
$D(H_r)$,  we have to first look for square integrable solutions of the
equations  \be H_r^* \phi_{\pm} = \pm i \phi_{\pm},
\label{def}
\ee where $H_r^*$ is the adjoint of $H_r$ (note that $H_r^*$ is given by the
same differential operator as $H_r$ although their domains might be
different).  Let $n_+(n_-)$ be the total number of square-integrable,
independent solutions  of (\ref{def}) with the upper (lower) sign in the right
hand side. The quantities $n_\pm$ are called the deficiency indices of $H_r$. Now
$H_r$ falls in one of the following categories  :\\ 1) $H_r$ is (essentially)
self-adjoint iff $( n_+ , n_- ) = (0,0)$.\\ 2) $H_r$ is not self-adjoint in $D(H_r)$ but admits self-adjoint extensions iff $n_+ = n_- \neq 0$.\\ 3)  $H_r$ has no self-adjoint extensions if $n_+ \neq n_-$.\\

To proceed, we note from eqn. (\ref{coupling}) that for any non-zero value of the dipole
moment, the parameter $\lambda < 0$. We restrict our analysis to the range $-\frac{1}{4} \leq \lambda < 0$, as the analysis for bound states in polar molecules with $\lambda < -\frac{1}{4}$ already exists in the literature \cite{leblond,case}. In terms of the variable 
$\nu=\sqrt{\lambda + \frac{1}{4}}$, the solutions of eqn. (\ref{def}) which are square integrable
at infinity are given by
\begin{eqnarray}
\nonumber \phi_+ (r) = r^{\frac{1}{2}}H^{(1)}_\nu (re^{i \frac{ \pi}{4}}),\\
\phi_- (r) = r^{\frac{1}{2}}H^{(2)}_\nu (re^{-i \frac{ \pi}{4}}),
\end{eqnarray}
where $H_\nu$'s are Hankel functions \cite{abr}. 
For the moment we consider the case $\lambda \neq -\frac{1}{4}$, i.e. $ \nu \neq 0$.
The functions $\phi_{\pm}$ are bounded as $r \rightarrow \infty $. When $r
\rightarrow 0$, they behave as
\begin{eqnarray}
\nonumber \phi_+(r) \rightarrow \mathcal{C}_1(\nu)r^{\nu + \frac{1}{2}} +
\mathcal{C}_2(\nu)r^{-\nu + \frac{1}{2}},\\ \phi_-(r) \rightarrow
\mathcal{C}^*_1(\nu)r^{\nu + \frac{1}{2}} + \mathcal{C}^*_2(\nu)r^{-\nu +
\frac{1}{2}},
\end{eqnarray}
where $\mathcal{C}_1(\nu)=\frac{i}{\sin\nu \pi}\frac{1}{2^\nu} \frac{ e^{ - i
 \frac{3 \nu \pi}{4}}}{\Gamma (1 + \nu)}$,
 $\mathcal{C}_2(\nu)=-\frac{i}{\sin\nu \pi}\frac{1}{2^{-\nu}} \frac{ e^{ - i
 \frac{\nu \pi}{4}}}{\Gamma (1 -\nu)}$ and $\mathcal{C}_1^*(\nu)$ and
 $\mathcal{C}_2^*(\nu)$ are complex conjugates of $\mathcal{C}_1(\nu)$ and
 $\mathcal{C}_2(\nu)$ respectively. We see that $\phi_{\pm}$ are not square
 integrable near the origin when ${\nu}^2 \geq 1$.  In this case, $n_+ = n_- = 0$ and  
$H_r$ is (essentially) self-adjoint in the domain
 $D(H_r)$ \cite{reed}. On the other hand, both $\phi_{\pm}$ are square
 integrable when either $ -1 < \nu < 0$ or $ 0 < \nu < 1$. We therefore see
 that for any value of $\nu$ in these ranges, $H_r$ has deficiency indices $
 (1,1)$. In this case, $H_r$ is not self-adjoint on the domain $D(H_r)$ but admits
 self-adjoint extensions. The domain $D_\omega(H_r)$ in which $H_r$ is
 self-adjoint contains all the elements of $D(H_r)$ together with elements of
 the form $\phi_+ + {\mathrm e}^{i\omega} \phi_-$, where $ \omega \in R$ (mod
 $2 \pi$) \cite{reed}.  We now proceed to obtain the spectrum of $H_r$ in the
 domain $D_\omega(H_r)$.

The solution of the differential equation Eq. (\ref{radial})  can be written as
\begin{equation}
R(r) = B r^{\frac{1}{2}} H^{(1)}_\nu(qr)\,, \label{radial1}
\end{equation}
where $q^2=\epsilon$. Note that in the limit $r \rightarrow 0$,
\begin{eqnarray}
\nonumber \phi_{+}(r) + e^{i \omega} \phi_{-}(r) \rightarrow
 \left[\mathcal{C}_1(\nu) +e^{i\omega}\mathcal{C}^*_1(\nu)\right]r^{\nu +
 \frac{1}{2}}\\
 +\left[\mathcal{C}_2(\nu)+e^{i\omega}\mathcal{C}^*_2(\nu)\right] r^{-\nu +
 \frac{1}{2}} \label{match1}
\end{eqnarray}
and
\begin{equation}
R(r) \rightarrow \mathcal{D}_1(\nu,q)r^{\nu + \frac{1}{2}} +
\mathcal{D}_2(\nu,q)r^{-\nu + \frac{1}{2}}, \label{match2}
\end{equation}
where $\mathcal{D}_1(\nu,q)=
\frac{i}{\sin\pi\nu}\frac{e^{-i\pi\nu}q^\nu}{2^\nu\Gamma(1+\nu)}$
and $\mathcal{D}_2(\nu,q)=
-\frac{i}{\sin\pi\nu}\frac{q^{-\nu}}{2^{-\nu}\Gamma(1-\nu)}$. If
$R(r) \in D_\omega(H_r)$, then the coefficients  of $r^{\nu +
\frac{1}{2}}$ and $r^{- \nu + \frac{1}{2}}$ in Eq. (\ref{match1})
and (\ref{match2}) must match. Comparing these coefficients we get
the bound state energy as
\begin{equation}
 E= - \frac{\hbar^2}{2\mu} \left [ \cos\frac{\pi\nu}{2}+
    \cot(\frac{\omega}{2}+\frac{\pi\nu}{4})\sin\frac{\pi\nu}{2}\right ]
    ^{\frac{1}{\nu}}\,. \label{eigen1}
\end{equation}
 Thus we see that for a given value of $\nu$ within the allowed range, $H_r$
admits a single bound state with energy given by Eq. (\ref{eigen1}). It may be
noted that for a fixed $\nu$, the bound state exists only for those values of
$\omega$ such that the quantity in first bracket in Eq. (\ref{eigen1}) is
positive. From Eq. (\ref{radial1}) and keeping in mind that
$\epsilon$ is negative,  we see that the the  bound state eigenfunction
is given by
\begin{equation} R(r) = B r^{\frac{1}{2}}
H^{(1)}_\nu(i \sqrt{|\epsilon|} r),
\end{equation}
where $B$ is the normalization constant. The bound state energy and the
eigenfunction depends on the choice of the self-adjoint extension parameter
$\omega$, which classifies the inequivalent boundary conditions.

The case for $\nu = 0$ or $\lambda = - \frac{1}{4}$ can be handled in a
similar fashion. The bound state energy and the wave function in this case are
given by \be E = -\frac{\hbar^2}{2\mu} {\exp}\left[\frac{\pi}{2}  {\cot}
\frac{\omega}{2}\right]\,, \ee and \be \psi (x) =  \sqrt{-2 \epsilon x} K_0\left( \sqrt{-\epsilon}
x\right)\,. \ee respectively, where $K_0$ is the modified Bessel function \cite{abr}.

We now come to the important issue of critical dipole moment of
polar molecules, which is  required to bind an electron. The above
analysis shows that for $-1 < \nu < 1$, the radial Hamiltonian
describing an electron in the field of a polar molecule admits a
single bound state. This implies that if  constant
$\lambda$ is in the range $- 1/4\leq\lambda < 3/4 $, the system
admits a bound state. Now recall from eqn. (\ref{coupling}) that $\lambda$ must be negative.
Combining these, we can conclude that for any real value of
$\lambda$ such that $-\frac{1}{4} \leq \lambda < 0$, the system
describing an electron in a dipole field admits a single bound
state. This conclusion is remarkably different from the statement in
the literature that $\lambda$ must be less than $ -\frac{1}{4}$ for bound states to
exist \cite{leblond}. Thus the mathematical analysis suggests that any molecule with a non-zero but arbitrarily
small dipole moment may be able to capture an electron to form a
bound state. In particular, this argument shows that molecules such as
$H_2S$ and $HCl$, whose dipole moments are smaller than the
critical dipole moment obtained from the usual analysis can also
capture electrons. This is consistent with the experimentally
observed anomalous electron scattering in these molecules \cite{leblond,rohr}, which the
previous analysis \cite{leblond} was not able to account for.  Our
model also predicts the existence of a single bound state, which is
again in qualitative agreement with the experimental observations,
in contrast to the usual treatment which predicts an infinite number
of bound states \cite{case}.

The exact numerical value of the bound state energy would depend on
the choice of the self-adjoint extension parameter $\omega$ which
characterizes the boundary conditions at the origin.
\begin{figure}
\includegraphics[width=0.4\textwidth, height=0.2\textheight]{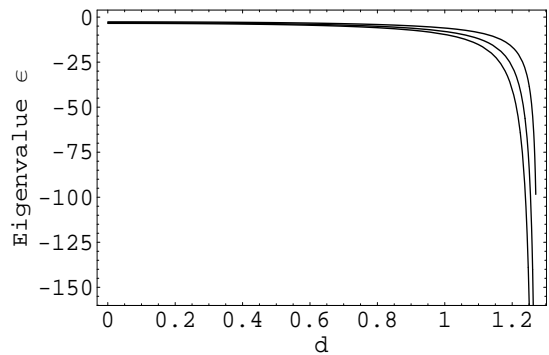}
\caption {A plot of binding energy of electron as a function of 
$d = \frac{2\mu e D}{\hbar^2}$ of the polar molecules for three different values of the self-adjoint
extension parameter $\omega$. From top to bottom $\omega= \frac{\pi}{10}, \frac{\pi}{8},\frac{\pi}{6}$ respectively.}
\end{figure}
Taking into account Eq. (\ref{coupling}), Eq.(\ref{eigen1}) and the condition
$-\frac{1}{4}\leq\lambda<0$, we have plotted the binding energy as a function of the dipole
moment $d$ in FIG. 1. It is clear that molecules with arbitrary small dipole
moment can bind electron, and that the bound state energies may also be very
small.

We now address the issue of breaking of scale invariance due to
quantization in this system, leading to a quantum mechanical anomaly. This feature is crucial for the formation of any bound state in this system, which is classically scale invariant. The relevance of quantum anomaly for the formation of bound states in polar molecules, with dipole moments greater than $D_0$, has been discussed in the literature \cite{camblong1}. Our analysis of the scaling anomaly applies even when the dipole moment is less than $D_0$.
Consider the case for $\nu \neq 0$. The anomaly arises as the scaling operator
$\Lambda = \frac{-i}{2} (r \frac{d}{dr} + \frac{d}{dr} r)$ acting on an
arbitrary element $\phi \in D_\omega(H_r)$, takes the wavefunction out of the
domain of the Hamiltonian. This can be seen as follows. In the limit $r
\rightarrow 0$, we have
\begin{eqnarray}
\nonumber \Lambda \phi(r)  \rightarrow\frac{(1+\nu)}{i}
\left[\mathcal{C}_1(\nu) +e^{i\omega}\mathcal{C}^*_1(\nu)\right]r^{\nu +
\frac{1}{2}}\\
+\frac{(1-\nu)}{i}\left[\mathcal{C}_2(\nu)+e^{i\omega}\mathcal{C}^*_2(\nu)\right]
r^{-\nu + \frac{1}{2}} \label{scale1}
\end{eqnarray}
In order for $\Lambda \phi(r) \in D_\omega(H_r)$, we must have $\Lambda \phi(r)
\sim C \phi(r)$ where $C$ is a constant. However, the two terms on the
r.h.s. of Eq. (\ref{scale1}) are multiplied by two different factors ,
i.e. $(1 + \nu )$ and $(1 - \nu )$. Due to the presence of these different
multiplying factors, we see that $\Lambda \phi(r)$ in general does not belong
to $D_\omega(H_r)$. Scale invariance is thus broken at the quantum level for
generic values of $\omega$, due to the choice of the domains of self-adjointness. Since the domains encode the boundary conditions, which in turn capture the effects of the short-range interactions in the polar molecules, we can qualitatively say that the short distance physics is responsible for breaking the scale invariance. 
However, from Eq. (\ref{scale1}) it is clear that
for special choice of $\omega = -\frac{ \nu \pi}{2}$ and
$ \omega = -\frac{ 3 \nu \pi}{2}$,
$\Lambda \phi(r) \in
D_\omega(H_r)$ and the scaling symmetry is recovered \cite{kumar1,kumar2,kumar3}.
For these choices of $\omega$, the bound states do not exist. A similar analysis can be performed when $\nu=0$ as well.

Finally, we would like to mention that the usual analysis assuming $\lambda < - \frac{1}{4}$ produces an unphysical Hamiltonian which is unbounded from below. This is not to be taken too seriously as the short distance physics would certainly cure this problem. Again, without any detailed knowledge of the short distance physics, and with $\lambda < - \frac{1}{4}$, it is possible to use renormalization group techniques to analyze this problem with a cutoff in the radial variable $r$ \cite{rajeev}. In the limit of the vanishing cutoff, such an analysis produces a beta function for the coupling $\lambda$, which indicates an ultraviolet stable fixed point at $\lambda = -\frac{1}{4}$. Thus a renormalization group analysis of the strong coupling regime indicates that the coupling in the parameter space would flow to a value $\lambda = -\frac{1}{4}$. 

In conclusion, here we have shown that it is possible for polar molecules to form bound state with electrons even though their dipole moments are arbitrarily small. This is possible through proper choice of domains which makes the corresponding Hamiltonian self-adjoint. In our interpretation, the domains or equivalently the boundary conditions obtained using von Neumann's theory of self-adjoint extensions capture some of the short distance physics, which in turn leads to a quantum mechanical scaling anomaly thereby producing the bound state. Our approach is consistent with the experimentally observed anomalous electron scattering by molecules such as 
$H_2S$ and $HCl$, which could not be explained by the critical dipole moment obtained in ref. \cite{leblond}. Moreover, our method predicts a single bound state for each molecule, which is also in qualitative agreement with experimental observations. The detection of such anions in systems with low value of dipole moments would be an experimental challenge, as the binding energies in such systems are usually very small.

\vskip 0.5 cm

\noindent
{\bf Acknowledgment}\\

K.S.G. would like to thank A. Mullin for providing her Ph.D. thesis and for helpful comments. This work was done within the framework of the Indo-Croatian Joint Programme of Cooperation in Science and Technology sponsored by the Department of Science and Technology, India (DST/INT/CROATIA/P-4/05), and the Ministry of Science, Education and Sports, Republic of Croatia.

\end{document}